\documentclass[french,a4paper]{article-hermes}

\usepackage{url}
\usepackage{lgrind}

\newcommand{\Nom}{ Minerve }

\begin{document}



\proceedings{DECOR'2004, Déploiement et (Re)Configuration de Logiciels}{65}
\title[Des correctifs de sécurité à la mise à jour%
 ]%
      {%
Des correctifs de sécurité à la mise à jour 
}

\subtitle{%
Audit, déploiement distribué et injection à chaud}
\author{Nicolas Loriant - Marc Ségura-Devillechaise - Jean-Marc Menaud }

\address{%
\'Ecole des Mines de Nantes, projet Obasco EMN/INRIA, \\
\{nloriant,msegura,jmenaud\}@emn.fr \hfill \url{http://www.emn.fr/x-info/arachne}}

\resume{La complexité toujours croissante des logiciels implique un accroissement des bogues. Ces derniers restent le 
 vecteur principal d'attaque des pirates informatiques et imposent aux éditeurs logiciels d'émettre régulièrement des mises à jour pour leurs corrections. Cependant, peu d'utilisateurs sont prêts à arrêter leurs applications et à investir le temps nécessaire pour mettre  à jour leur poste de travail. Dans les faits, les correctifs sont rarement déployés, même quand ceux-ci sont critiques. \\
Nous proposons un atelier de déploiement permettant à un administrateur système d'appliquer à chaud, à distance et sans l'intervention des utilisateurs, des rustines de sécurité. Pour éviter l'arrêt des logiciels concernés, notre approche est basée sur un tisseur dynamique d'aspects, Arachne. Notre second outil, Minerve, s'intègre dans le processus de mise à jour standard : il prend en entrée une rustine source produite par \texttt{diff} et génère, quand cela est possible, une rustine dynamique déployable à la volée. Il permet de consulter les rustines exprimées dans un langage dédié et facilite ainsi la tâche d'audit des mises à jour déployées.}

\abstract{The ever growing software complexity suggests that they will never be bugfree and therefore secure. Software compagnies regulary publish updates. But maybe because of lack of time or care or maybe because stopping application is annoying, such updates are rarely if ever deployed on users' machines. \\
We propose an integrated tool allowing system administrators to deploy critical security updates on the fly on applications running remotly without end-user intervention. Our approach is based on an aspect weaving system, Arachne, that dynamicaly rewrites binary code. Hence updated applications are still running while they are updated. Our second tool Minerve integrates Arachne within the standart updating process: Minerve takes a patch produced by \texttt{diff} and eventually builds a dynamic patch that can later be woven to update the application on the fly. In addition, Minerve allows to consult patches translated in a dedicated language and hence eases auditing tasks.}

\motscles{mise à jour de programme à chaud, programmation par aspect, déploiement distribué, sécurité}


\maketitlepage

\section{Introduction}

\indent Pour l'année 2003, 80\% des attaques informatiques subies par les entreprises sont liées à un manquement dans le déploiement des correctifs de sécurités\cite{2004:devoteam:enquete}. En effet, la plupart des attaques informatiques exploitent des failles de sécurité répertoriées. Dans le cas de la dernière attaque médiatisée, le virus Sasser, une rustine corrigeant la vulnérabilité exploitée était disponible quinze jours avant la propagation du virus. De ce fait, une grande partie des attaques informatiques subies pourraient être évitées en menant de front à la fois, une veille attentive sur les failles de sécurité détectés par les organismes spécialisés, et par une application rapide des correctifs sur l'ensemble des machines concernées. Cette double mission est généralement octroyée à un administrateur système promu au rang de RSSI (Responsable de la Sécurité du Système Informatique). En l'absence d'outils adéquats, le RSSI est toutefois dénué de moyens pour lutter efficacement contre les d'attaques potentielles. En effet, la simple mission de veille lui demandera un peu plus de 13 semaines de travail à raison de cinq minutes pour chacune des 5500 alertes annuelles recensées\cite{2003:cert:overview}. En supposant qu'il ait la charge d'administrer 100 machines, qu'elles ne soient concernées que par 1\% des  alertes et que l'application d'un  correctif demande une heure de travail, le temps nécessaire à leur déploiement s'élèvera au total à plus de 157 semaines de travail par an.\\
\indent Nous proposons dans cet article une approche semi-automatique pour le déploiement des correctifs de sécurités. Le but de cette approche est de réduire significativement le temps nécessaire au déploiement des correctifs. Le principal problème n'est pas tant l'application du correctif en lui même mais d'obtenir de l'utilisateur l'accès aux machines nécessaires et l'arrêt des applications à mettre à jour (prise de contact, sauvegarde de la session et des documents etc..). 
Notre approche est  basée sur deux outils, Minerve et Arachne. Le premier réduit le temps de certification des correctifs par le RSSI. Le second permet le déploiement des correctifs sans avoir à arrêter l'application à mettre à jour. La mise à jour à chaud évite au RSSI d'obtenir l'accord de l'utilisateur pour appliquer un correctif et réduit ainsi significativement leur temps de déploiement. La section \ref{sec:arch} présente une vue d'ensemble de notre atelier de déploiement et son intégration dans le processus traditionnel de mise à jour des logiciels libres. Les sections \ref{sec:etude} et \ref{sec:ara} présentent respectivement nos outils : Minerve et Arachne. La section \ref{sec:eval} décrit une évaluation de l'atelier appliqué aux derniers trous de sécurité. Avant de conclure, la section \ref{sec:art} résume quelques  principaux travaux connexes.\\
\section{Atelier de déploiement}\label{sec:arch}
\indent Traditionnellement dans les logiciels libres, lorsqu'une une faille de sécurité est découverte, un correctif est proposé sous la forme d'une rustine source (patch) produite par l'utilitaire {\tt diff}. Cet outil permet de lister les différences textuelles entre deux fichiers sources.\\
\indent Lorsque le RSSI prend connaissance d'une faille de sécurité qui affecte son parc informatique, il vérifie en examinant les sources du logiciel que le correctif ne remet pas en cause son intégrité. Il doit alors recompiler, redéployer le logiciel corrigé sur l'ensemble du parc puis ré-initialiser le logiciel et/ou la machine. \\
\indent Ce processus de mise à jour présente plusieurs lacunes. En premier lieu, l'audit des rustines fournies par les éditeurs logiciels est difficile à réaliser car les fichiers de différences textuelles entre les versions d'un logiciel ne fournissent pas d'informations pertinentes et claires des modifications apportées par la rustine. En second lieu, les failles de sécurité sont toujours présentes tant que le logiciel incriminé n'a pas été stoppé. Or pour certaines applications comme les serveurs de vente en ligne, cette contrainte n'est pas acceptable. De plus, un RSSI ne peut prendre la responsabilité d'arrêter l'application d'un utilisateur sans savoir si ce dernier a préalablement sauvegardé son travail. En dernier lieu, le déploiement du logiciel corrigé doit se faire machine après machine ce qui, sans outil adapté, est une perte de temps importante et peut être compliqué dans un environnement mobile. \\
\indent Notre atelier logiciel est destiné à faciliter la lecture des correctifs, à automatiser leurs déploiements et enfin à les appliquer dynamiquement sans arrêt des logiciels. Dans ce dessein, notre atelier est basé sur deux outils : Minerve et Arachne. \\
\indent Le premier est chargé à la fois de valider le correctif dans le cadre d'une application dynamique et de transformer les rustines sources très peu lisibles en fichiers de description des modifications présentant au RSSI une vue synthétique des modifications qui seront apportées. Minerve permet de présenter au RSSI le corps des fonctions avant et après modification en utilisant une symbolique pour désigner les points de changement. Après validation du RSSI en fonction des informations fournies par Minerve sur l'applicabilité dynamique du correctif, le fichier est compilé en binaire sous forme d'une librairie, puis déployé sur l'ensemble des machines concernées par la mise à jour. \\
\indent Le second, préinstallé sur l'ensemble du parc, est alors chargé d'appliquer le correctif. Sa principale fonction consiste à s'insérer dynamiquement et sans l'arrêter dans le logiciel incriminé, puis de récrire le code binaire du logiciel en vue de remplacer les fonctions impliquées dans la faille de sécurité par leurs nouvelles versions contenues dans la DLL précédemment diffusée. \\
\indent Finalement, le correctif n'est appliqué que pour la durée d'exécution du logiciel. En effet, en parallèle à cette opération, une mise à jour classique par application du patch sur les sources est effectuée. Les prochaines exécutions du logiciel seront donc liées à la nouvelle version \cite{2004:novadigm:hp}. Dans notre atelier, le travail de l'administrateur se limite à effectuer une veille sur les failles et correctifs diffusés, valider les correctifs concernés par les machines/logiciels qu'il administre et finalement indiquer à nos outils le sous ensemble des machines/logiciels à mettre à jour.

\section{Minerve : vérification de l'applicabilité dynamique des correctifs}\label{sec:etude}

\indent Les rustines sources fournies par les éditeurs logiciels sont générées et déployées à partir des programmes {\tt diff} et {\tt patch}. Ces outils manipulent uniquement des différences textuelles entre fichiers sources sans exploiter des informations sémantiques sur les modifications apportées. Si cette approche permet de valider l'application d'un correctif dans un cadre statique (ou off line), elle n'est pas suffisante dans un cadre dynamique (ou on line). Minerve est chargé de classifier les changements effectués dans le code de l'applicatif puis d'informer le RSSI de l'applicabilité dynamique de ceux-ci (section 3.1). Enfin, un fichier des modifications plus expressif qu'un simple {\tt diff} est produit pour faire appel à l'audit du RSSI (section 3.2). \\

\indent En vue de démontrer la faisabilité de notre approche nous nous sommes limité à ne considérer que les programmes écrits en langage C ANSI et compilé sur les plateformes IA-32 dans un environnement Linux. Nous supposons également que les rustines sources fournies par les éditeurs sont cohérentes (i.e. que les sources corrigées passent la phase de compilation et fournissent un programme valide).

\subsection{Classification et Vérification}
\indent Le premier objectif de l'outil est de classifier les modifications à effectuer sur un programme cible écrit en C.
Le langage C est un langage procédural typé à effet de bord. Il contient deux sortes de types, les types simples qui reflètent les entités que le processeur peut manipuler efficacement, et les types structurés, construits par le développeur par agrégation de types simples. Une rustine source modifie le comportement d'un programme en modifiant la définition des procédures qu'il contient, et/ou en modifiant les types associés aux variables qu'il utilise. Si d'un point de vue statique les deux types  de modifications peuvent être traitées symétriquement, d'un point de vue dynamique, chacun des cas doit être traité différemment. 

\subsubsection{Modification de fonctions}\label{sec:remp_fct}

\indent Nous distinguons trois types de correctifs liés aux fonctions : l'ajout, la suppression et la modification. En réalité ces trois types de correctifs peuvent se résumer en un seul : la modification d'une fonction. 
En effet, l'ajout d'une fonction $g$ dans un programme implique qu'il existe au moins une fonction $f$ qui a été modifiée de sorte qu'elle utilise $g$. Dans le cas contraire l'ajout de $g$ n'aurait aucun sens. De même, la suppression d'une fonction $g$ nécessite qu'au moins une fonction $f$ soit modifiée. Dans le cas contraire, une erreur à la compilation du programme corrigé serait générée. Or, nous avons précisé que les rustines considérées étaient cohérentes. Nous nous limitons donc aux cas de la modification d'une fonction $ f $ et parlerons de $ f'$ pour désigner la nouvelle version de $ f$ (donc après application de la rustine). \\
\indent La modification d'une fonction peut entraîner le changement de sa signature et/ou le changement de son corps. Le changement de la signature d'une fonction est équivalent à l'ajout d'une nouvelle fonction. Pour le deuxième cas, le remplacement de $ f $ par $ f'$ sans modification de prototype, il est nécessaire de s'assurer que $f$ n'est et ne sera jamais plus exécutée \cite{1993:segal:podus,1998:gupta:osvc}. Cette première condition ne peut être vérifiée que dynamiquement par inspection de la pile d'exécution. Pour cela, nous nous appuyons sur l'outil Arachne présenté en section \ref{sec:ara}. De plus, le code ajouté (si il y a lieu) dans  $ f$ ne doit pas lire les données produites par $f$. Pour cela nous pensons utiliser un outil de propagation de type nommé Lackwit \cite{1997:lackwit:callahan}.

\subsubsection{Modification d'une variable}
\indent Nous distinguons deux types de modifications, celles liées au changement de type de base, et celles liées à la modification des types structurés.
\begin{itemize}
\item La redéfinition du type d'une variable nécessite deux opérations : la propagation de l'état de la variable et les modifications à apporter au code qui la manipule. Augmenter la taille d'une variable sans modifier son type numérique (ex: {\it int}$\rightarrow$ {\it long int}) ne pose pas de problème de conversion, puisque le nouveau type englobe l'ancien. Au contraire, la diminution de la taille d'une variable (ex: {\it long int}$\rightarrow$ {\it int}) ou la modification du type numérique de celle-ci (ex: {\it float}$\rightarrow$ {\it int}) n'est possible que si sa valeur courante peut-être contenue par le nouveau type ou si l'on dispose d'une fonction de conversion. Modifier la taille ou le type numérique d'une variable nécessite de modifier également le code binaire manipulant celle-ci. En effet, les instructions processeur, les registres, les drapeaux ainsi que les exceptions associées à des types différents sont distincts. La modification d'une instruction peut nécessiter de réorganiser les instructions voisines. Cette réorganisation peut s'étendre au remplacement complet de la fonction accédant à la variable dont le type a été modifié. Les registres globaux étant de la responsabilité des fonctions appelantes et les autres registres des fonctions appelées 
, la modification du type numérique d'une variable peut alors également nécessiter la modification des fonctions appelant les fonctions qui accèdent aux variables modifiées. Ce cas n'est néanmoins pas un problème puisqu'il est résolu par la technique d'instrumentation d'Arachne.
\item Modification de la définition d'un type structuré : nous avons choisi de présenter l'ajout d'un champ dans un type structuré puisqu'il illustre bien les problèmes liés à ce type de modifications (suppression, remplacement, inversion). La modification d'un type structuré peut modifier la contrainte d'alignement sur la représentation mémoire des instances de ce type. 
Certaines instructions de la plate-forme IA-32 peuvent avoir des comportement différents, voir ne pas fonctionner lorsque les données qu'elles manipulent ne sont pas alignées 
. Modifier toutes les variables existantes du type modifié imposerait alors également de mettre à jour l'ensemble des instructions manipulant ce type dans le programme d'origine. Or, la cohérence de la mise à jour ne pourrait alors être garantie qu'en stoppant le programme. Toutefois, une technique de réécriture appropriée permet de résoudre ce problème. En effet, un champ ajouté dans un type structuré lors d'une mise à jour n'existe pas dans le programme de base, seules les instructions ajoutées peuvent manipuler celui-ci. Il est ainsi possible de modifier ces dernières de façon à associer via une fonction de hachage, à chaque variable du type d'origine une variable représentant le nouveau champ. La création du nouveau champ peut se faire de manière paresseuse pour chaque variable. Ceci nous permet d'éviter la modification de toutes les variables de ce type lors de la mise à jour.
\end{itemize}
\indent Sachant que le RSSI est placé au centre de notre architecture, l'ensemble des informations de vérifications effectuées par Minerve ne lui sont présentées qu'à titre indicatif. Libre à lui de valider ou non la rustine dynamique. Pour l'aider dans cette tâche Minerve produit un fichier concentrant l'ensemble des modifications à apporter au logiciel incriminé dans un format lisible et compatible avec notre outil de transformation de code à chaud. La section suivante décrit le processus d'inspection de la rustine.
\subsection{L'audit du RSSI}\label{sec.arachne.audit}
\indent L'inspection du code de la rustine répond à deux buts. D'abord elle permet de s'assurer que la faille de sécurité recensée est véritablement corrigée. Puis, elle veille à ce que de nouveaux trous de sécurité ne soit pas introduits par erreur. Il peut également être utile d'adapter un correctif de sécurité au système local d'information. Par exemple, différents experts en sécurité préconisent que la correction d'une vulnérabilité s'accompagne de la mise en place d'une alarme signalant toute tentative d'exploitation de la vulnérabilité colmatée \cite{Sch00}. Une fois la rustine traduite par  \Nom dans le langage d'Arachne, le RSSI peut facilement l'éditer au cours de l'audit afin d'y placer une alarme qui va par exemple, déclencher le système de détection d'intrusion (IDS). \\
\indent Au contraire de \texttt{diff} et \texttt{patch} qui fonctionnent par substitution de lignes, \Nom utilise Arachne pour remplacer des fonctions aux comportements erronés par de nouvelles fonctions et redéfinir les accès fautifs aux variables globales. Une fois la rustine traduite par Minerve dans le langage d'Arachne, la rustine contient l'ensemble des entités (variables globales et fonctions) à redéfinir. Chaque rustine de sécurité est ainsi traitée comme une préoccupation à part, inspectable et éditable isolément des autres. De plus, le format de fichier est directement compatible avec Arachne \cite{ segura-menaud-al.cfse2003}, un tisseur dynamique d'aspect pour le langage C. \\
\indent Avec Arachne, chaque substitution ou \emph{aspect} est définie par un couple (coupe, action). Une coupe spécifie quand dans le flot d'exécution du programme, le code décrit dans l'action doit s'exécuter. Une coupe est une séquence d'appels de fonctions éventuellement terminée par un accès en lecture ou en écriture à une variable globale. Tous les éléments, sauf le dernier de la séquence, déterminent le contexte d'exécution de la coupe. Dans ce cadre d'étude, nous utilisons Arachne comme un outil d'instrumentation à chaud nous permettant d'intercepter tout accès à une fonction ou à une variable particulière. De ce fait, les abstractions langage de haut niveau propre à la programmation par aspect et fournies par Arachne ne seront pas utilisées dans notre solution. 

\section{Arachne : injection dynamique des correctifs}\label{sec:ara}
\indent Cette section décrit  les outils associés à Arachne permettant au RSSI de compiler la rustine dynamique avant de l'injecter dans l'application à mettre à jour, puis s'achève par une vue d'ensemble des mécanismes exploités dans la mise en \oe uvre d'Arachne.

\subsection{Les outils de compilation et de déploiement}\label{sec.arachne.compil}
\indent Arachne s'accompagne de deux outils, un compilateur d'aspects \texttt{acc}, générant une DLL native et un tisseur \texttt{weave} permettant d'injecter, \emph{tisser} les aspects compilés dans l'application incriminée. Indépendamment de notre hypothèse de travail qui suppose que la rustine traduite par \Nom est correcte, la compilation effectuée par \texttt{acc} constitue une vérification globale de la cohérence de la rustine. Plus précisément, elle garantit l'absence d'erreurs syntaxiques. De plus, à l'injection, \texttt{weave} vérifie que l'application à mettre à jour contient bien les symboles attendus par la rustine. Ainsi, si une rustine essaie de remplacer une fonction ou une variable globale qui n'est pas définie dans l'application, l'injection de la rustine échoue. Bien que ce type de vérification ne permette pas de garantir que la rustine corrige le trou de sécurité, il permet néanmoins d'assurer un niveau minimal de cohérence. Si la rustine nécessite plusieurs points d'instrumentation, Arachne garantit l'atomicité de la réécriture : soit toutes les fonctions/variables sont remplacées soit aucune. Enfin, une option particulière permet à Arachne de vérifier si une fonction à remplacer est en cours d'exécution et dans l'affirmative, d'attendre la fin d'exécution avant le tissage.

\subsection{Implémentation d'Arachne}\label{sec.arachne.impl}
\indent Le tisseur d'Arachne, piloté par la commande \emph{weave}, récrit à la volée le code binaire de l'application dans laquelle il doit injecter les aspects \cite{segura-menaud-al.aosd2003}. Il est difficile de couvrir les différentes stratégies de réécriture utilisées par le tisseur. Nous précisons ici toutefois le principe général qui permet à Arachne de remplacer une fonction par une autre avant de préciser les problèmes techniques qui compliquent sa mise en oeuvre. \\

\indent Sur une machine Pentium sous Linux, un appel de fonction au niveau source se traduit dans le code binaire de l'application par une instruction assembleur \texttt{CALL}  suivie de l'adresse de la fonction à invoquer. Arachne désassemble le code binaire de l'application et y recherche les instructions \texttt{CALL}. Pour cela, il utilise les tables de symboles générées à la compilation pour retrouver, à partir de l'adresse manipulée par l'instruction \texttt{CALL}, le nom symbolique de la fonction invoquée. Au tissage, Arachne charge la DLL contenant les aspects et récrit les différents \texttt{CALL} vers la fonction à remplacer par des \texttt{CALL} vers la fonction appropriée de la DLL. Un mécanisme similaire mais plus complexe permet à Arachne de remplacer les lectures/écritures sur les variables globales. \\
\indent En pratique, plusieurs problèmes compliquent la mise en oeuvre du principe de réécriture décrit plus haut. En résumé, il s'agit de garantir l'atomicité des réécritures effectuées dans le code objet de l'application, d'assurer la cohérence de l'exécution, de garantir que l'ensemble des aspects de la DLL soit tissés d'une manière atomique (c'est-à-dire qu'un aspect ne puisse pas s'exécuter alors que les autres sont en train d'être tissés), de gérer les contraintes imposées par le système d'exploitation en terme de séparation des espaces d'adressage et finalement de résoudre les problèmes d'efficacité. Arachne résous les problèmes d'atomicité des réécritures par un système de verrous tournants reposant sur la taille des instructions de sauts courts et sur la taille des accès mémoire atomiques sous Pentium. La cohérence de l'exécution est assurée par un crochet généré à la volée lors du tissage : Ce crochet sauvegarde et restaure les registres nécessaires. L'utilisation d'un test sur une garde garantit que l'ensemble des aspects de la DLL sont tissés de manière atomique. La séparation imposée des espaces d'adressage par le système d'exploitation empêche un processus de modifier l'espace mémoire d'un processus tiers. La première fois qu'Arachne tisse dans un processus, il accède à son espace mémoire en utilisant les fonctions de déverminage \texttt{ptrace}, puis il créé un processus léger d'instrumentation partageant le même espace d'adressage que l'application. Arachne pilote ensuite ce processus léger d'instrumentation en utilisant une interface d'interconnexion : le tissage s'effectue ainsi globalement sans interruption de l'application. Arachne adresse les problèmes d'efficacité par des caches conçus pour n'examiner une fois au plus le code binaire de l'application.

\section{\'Evaluation}\label{sec:eval}

\indent Cette section présente une évaluation de notre atelier composé de Minerve et Arachne. Pour cela, nous avons appliqué notre méthodologie aux mises à jour de sécurité concernant des logiciels libres, publiées depuis le début de l'année 2002 par le CERT/CC. Après un présentation de l'organisme et de l'échantillon d'étude, nous présentons nos résultats sur l'ensemble de l'échantillon puis décrirons un exemple concret de rustine dynamique.

\subsection{\'Echantillon d'étude}

\indent Le CERT/CC pour \og Computer Emergency Response Team / Coordination Center\fg \- est un centre d'expertise créé en novembre 1988, suite à l'apparition du ver \emph{Morris}. Par ses activités de veille, de diffusion, d'alerte et de formation, il cherche à améliorer la sécurité des systèmes informatiques. Sa longévité, son indépendance vis-à-vis des éditeurs de logiciels ainsi que son rôle préventif et curatif font de lui un acteur de référence mondiale du domaine. Depuis 1988, il a ainsi collecté une des bases de données les plus complète en matière de failles de sécurité. 
Toutes les failles relevées par le CERT/CC sont classées suivant trois catégories. La première, les alertes recensent les attaques perpétrées contre des systèmes informatiques. La seconde, les rapports de vulnérabilité listent les failles dans les applications. Enfin la dernière catégorie, les avis référencent un ou plusieurs rapports de vulnérabilité. Les avis se limitent aux failles de sécurité critiques sur des logiciels largement répandus et fournissent systématiquement les correctifs associés. \\
\indent Notre étude s'est donc portée sur l'ensemble des avis publiés par le CERT/CC depuis le début de l'année 2002 jusqu'à aujourd'hui. Cette période représente un total de 67 avis, 30\% concernent des produits Microsoft, 20\% des logiciels propriétaires autres que Microsoft, 40\% référencent des failles dans des logiciels libres et les 10\% restants concernent de l'informatique embarquée dans des matériels réseau. 14 avis concernent des logiciels libres sous Linux. Nous en avons écarté deux parce qu'ils affectaient des versions obsolètes introuvables. \\
\indent Nous avons constaté que \nombre{44\%} des failles examinées sont des débordements de tampon : Ce type de dysfonctionnement peut permettre à un utilisateur malveillant de faire exécuter du code arbitraire par le processus visé. \nombre{19\%} présentaient des erreurs de format : ces erreurs affectent les fonctions à nombre de paramètres  variables (ex : {\tt printf}). Un utilisateur malveillant peut exploiter cette faille pour obtenir des informations sur le processus visé. \nombre{12\%} des cas recensent des doubles désallocations mémoire. Cette faille provoque normalement une erreur de segmentation mais peut sous certaines conditions permettre l'exécution de code arbitraire. \nombre{6\% } des avis examinés présentent un débordement d'entier : Ceci se produit lorsque la valeur d'un entier dépasse la limite du type. Enfin, \nombre{19\% } des trous de sécurité résultent de combinaisons des erreurs de conceptions citées ci-dessus. \\
\indent Toutes ces failles se basent sur des hypothèses faites par le programmeur sur les paramètres du programme mis en défaut par un utilisateur malveillant. Ainsi, ces failles peuvent se corriger très simplement : il suffit d'ajouter des tests sur les valeurs des paramètres. L'examen des rustines correctives de ces trous de sécurité nous a permis de dégager le contenu des modifications apportées à un programme pour corriger une faille. Nous avons pu constater la corrélation avec le type des dysfonctionnements recensés, puisque \nombre{90\%} des correctifs source ne contenaient que des modifications de fonctions sans changement de signatures. Seulement \nombre{10\%} contenait à la fois des modifications de code et des transformations de types. \\
\indent Ce constat ainsi que les expérimentations que nous avons réalisées, nous permettent d'affirmer que notre atelier est applicable à tous les trous de sécurité étudiés. \\

\subsection{Un exemple de mise à jour}
\indent Nous présentons dans cette section une des failles que nous avons pu corriger automatiquement grâce à Minerve. Il s'agit d'une vulnérabilité découverte en juin 2002 et qui concerne le serveur de communication \textsc{OpenSSH} (avis numéro 18/2002 du CERT/CC) \cite{2003:cert:overview}. 
Cette faille concerne le mécanisme d'authentification du protocole SSH2 : {\tt challenge response}. Elle affecte les versions de {\tt 2.3.1p1} à {\tt 3.3} configurées avec les options {\tt SKEY} ou {\tt BSD\_AUTH}. C'est le cas pour l'installation par défaut dans la plupart des distributions Linux. Cette faille consiste en un débordement d'entier exploitable par des utilisateurs malveillants afin d'exécuter du code arbitraire par le programme résidant {\tt sshd}, exécuté généralement avec les droits super-utilisateur.

\subsubsection{Rustine source}
\indent La rustine source fournie par les développeurs de \textsc{OpenSSH} corrige les fonctions {\tt input\_\-userauth\_\-info\_\-response} et {\tt input\_\-userauth\_\-info\_\-response\_\-pam}. Les modifications se bornent à ajouter des tests de validité sur le paramètre {\tt nresp} dans ces deux fonctions. Lorsque ces tests de validité échouent, la fonction {\tt fatal} est déclenchée pour terminer la session en cours. Comme le montre la figure \ref{fig:patch}, l'examen de la rustine produite par \texttt{diff} ne permet pas de comprendre facilement les changements effectués par le correctif.\\
\lagrind[ht]{patchSource.tex}{Une partie du fichier de différences {\tt diff} de la mise à jour de sécurité de {\tt sshd} référencée par le CERT/CC par {\tt CA-2002-18}}{fig:patch}

\subsubsection{Rustine dynamique}
\indent Minerve produit à partir de la rustine source une rustine dynamique. La rustine dynamique contient essentiellement une suite d'aspects qui réalisent le remplacement des fonctions du programme de base,  {\tt input\_\-userauth\_\-info\_\-response} et {\tt input\_\-userauth\_\-info\_\-response\_\-pam}. Les fonctions corrigées possèdent le suffixe {\tt "\_new"}. Finalement, les anciennes fonctions pouvant être appelées dans {\tt sshd} indifféremment par un appel direct de fonction ou par un appel par pointeur, deux aspects sont nécessaires pour chacune de ces fonctions afin de réaliser leur remplacement à coup sûr. La figure \ref{code:arachne} illustre ces aspects pour la fonction {\tt input\_\-userauth\_\-info\_\-response}. \\
\indent La coupe de l'aspect {\tt ReplaceFunctionCall} capture l'ensemble des appels directs à cette fonction (ligne 2 du source figure \ref{code:arachne}). L'action remplace son exécution par celle de la nouvelle version à laquelle les même arguments sont fournis (ligne 3). Cet aspect permet ainsi de remplacer tous les appels directs à {\tt input\_\-userauth\_\-info\_\-response}. La coupe du second aspect {\tt ReplacePointer} capture les accès en lecture sur l'adresse de la fonction {\tt input\_\-userauth\_\-info\_\-response} (ligne 6). C'est cette valeur qui peut être ensuite utilisée dans un appel de fonction par pointeur. La valeur lue est remplacée dans l'action de l'aspect (ligne 7) par l'adresse de la nouvelle version de la fonction définie dans la rustine dynamique. Ainsi toutes les lectures de l'adresse de l'ancienne fonction seront remplacées par une lecture de l'adresse de la nouvelle fonction.

\lagrind[ht]{patch.tex}{Aspects générés par Minerve pour le remplacement systématique d'une fonction}{code:arachne}

\indent Comme le montre la figure \ref{code:arachne}, la syntaxe des rustines produites par Minerve facilite réellement le travail d'audit des correctifs. Sur l'ensemble des correctifs publiés par le CERT/CC en 2002, le gain de temps est réellement sensible. Si on exclut les temps réseau, dans nos expériences, une mise à jour est déployée en moyenne sur un poste en 253 $\mu s$ sans même que l'utilisateur en ait conscience. De plus avec notre atelier, le RSSI ne gère pas la mise à jour d'un poste mais de la totalité du parc. Le temps pour appliquer un correctif est indépendant du nombre de machines composant le parc.

\section{Travaux apparentés}\label{sec:art}
\indent \`A notre connaissance, aucune autre solution ne propose \emph{à la fois} la vérification de l'applicabilité d'une mise à jour dynamique et son déploiement à chaud. La vérification et la cohérence des mises à jour ont essentiellement été étudiées sous l'angle de contraintes sur  l'état de l'exécution de l'application à mettre à jour \cite{1996:tse:gupta} ou sur la structure \cite{1993:segal:podus} de son code. Notre approche se concentre elle sur la structure de la rustine source traditionnellement distribuée. Dyninst \cite{buck00api} et Vulcan \cite{edwards.srivastava.ea:vulcan}  explorent tout deux la possibilité de réécrire le code binaire à chaud. Contrairement à notre approche, ces outils travaillent au niveau de l'assembleur et ne garantissent pas l'atomicité de la réécriture. C'est un problème quand la mise à jour implique d'instrumenter le programme  à plusieurs endroits.

\section{Conclusion}\label{sec:con}
\indent Nous avons présenté dans cet article un atelier pour le déploiement à chaud de correctifs de sécurité. Cet atelier est composé de deux outils. Le premier, Minerve, assure la faisabilité d'un déploiement à chaud et réduit le temps nécessaire à la certification par le RSSI d'un correctif. Le second, Arachne, applique à chaud sur un programme en cours d'exécution, la rustine dynamique produite par Minerve. Notre atelier s'insère dans le processus traditionnel de mise à jour des logiciels libres basé sur les utilitaires \texttt{diff} et \texttt{patch}. L'utilisation de notre atelier permet un gain de temps significatif dans le déploiement et apporte donc une plus grande réactivité quant à la correction des failles. \\
\indent Concrètement, Minerve, transforme un correctif peu lisible en un fichier décrivant les modifications à apporter au logiciel. Ces modifications sont exprimées sous forme d'un aspect dont les coupes se limitent au remplacement de fonctions et/ou à l'accès à une variable. En ce sens nous n'utilisons pas l'ensemble du pouvoir d'expression de la programmation par aspect. Le second, Arachne, injecte dynamiquement l'aspect compilé dans l'application  à mettre à jour. Pour des raisons techniques, nous nous sommes pour l'instant limités aux programmes écrits en C s'exécutant sur une plate-forme IA32 dans un environnement Linux. L'atelier Minerve/Arachne a permis de déployer l'ensemble des correctifs de sécurité publiés par le CERT/CC sur les logiciels libres depuis 2002.

\bibliography{dcor}


\end{document}